\documentclass[12pt]{article}

\ifx\pdfoutput\undefined
\usepackage[dvips,bookmarks]{hyperref}
\else
\usepackage{hyperref}
\fi
\hypersetup{colorlinks=false,bookmarksopen,bookmarksnumbered,citecolor=blue,
   pdfstartview=FitH}

\usepackage[pdftex]{graphicx}	
\usepackage{latexsym}

\usepackage{amssymb,amsfonts,amsmath}
\usepackage{graphicx} 
\usepackage{indentfirst}

 \usepackage{bbm}

\parskip=\medskipamount

\newcommand {\cD}{{\cal D}}
\newcommand {\cE}{{\cal E}}

\newcommand {\cK}{{\cal K}}
\newcommand {\cL}{{\cal L}}
\newcommand {\cM}{{\cal M}}
\newcommand {\cN}{{\cal N}}

\def\a{\alpha}

\def\b{\beta}

\def\d{\delta}
\def\e{\epsilon}

\def\g{\gamma}
\def\G{\Gamma}

\def\k{\kappa}

\def\m{\mu}

\def\q{\theta}

\def\s{\sigma}

\def\x{\xi}

\def\D{\Delta}
\def\F{\Phi}
\def\J{\Psi}
\def\L{\Lambda}
\def\O{\Omega}
\def\P{\Pi}
\def\Q{\Theta}
\def\S{\Sigma}
\def\U{\Upsilon}
\def\X{\Xi}

\def\rd{{\rm d}}
\def\ri{{\rm i}}

\newcommand{\ad}{{\dot{\alpha}}}                           
\newcommand{\bd}{{\dot{\beta}}}                            
\newcommand{\ve}{\varepsilon}                            
\newcommand{\cDB}{{\bar\cD}}                            
\newcommand{\DB}{\bar{D}}

\newcommand{\pa}{\partial}                           
\newcommand{\hf}{\frac12}

\newcommand{\be}{\begin{equation}}
\newcommand{\ee}{\end{equation}}
\newcommand{\bea}{\begin{eqnarray}}
\newcommand{\eea}{\end{eqnarray}}
\newcommand{\non}{\nonumber}
\def\double #1{#1{\hbox{\kern-2pt $#1$}}}

\newcommand{\gd}{{\dot\g}}
\newcommand{\dd}{{\dot\d}}

\newcommand{\ts}{{\tilde{\s}}}

\newcommand{\teb}{{\bar{\theta}}}
\newcommand{\qb}{{\bar{\theta}}}

\newcommand{\Fb}{{\bar{\F}}}

\newif\ifdtup

\newcommand{\bsubeq}{\begin{subequations}}
\newcommand{\esubeq}{\end{subequations}}

\numberwithin{equation}{section}

\begin{document}
\begin{titlepage}
\begin{flushright}
October, 2017\\
\end{flushright}
\vspace{5mm}

\begin{center}
{\Large \bf  
Complex three-form supergravity and membranes
}\\ 
\end{center}

\begin{center}

{\bf 
Sergei M. Kuzenko${}^{a}$ 
and
Gabriele Tartaglino-Mazzucchelli${}^{b}$
} \\
\vspace{5mm}

\footnotesize{
${}^{a}${\it School of Physics and Astrophysics M013, The University of Western Australia\\
35 Stirling Highway, Crawley W.A. 6009, Australia}}  
~\\
\vspace{2mm}
\texttt{sergei.kuzenko@uwa.edu.au
} 
~\\
\vspace{2mm}
\footnotesize{
${}^{b}${\it Instituut voor Theoretische Fysica, KU Leuven,\\
Celestijnenlaan 200D, B-3001 Leuven, Belgium}
}
\vspace{2mm}
~\\
\texttt{
gabriele.tartaglino-mazzucchelli@kuleuven.be}\\
\vspace{2mm}

\end{center}

\begin{abstract}
\baselineskip=14pt

There exist two variants of the old minimal formulation for $\cN=1$ supergravity
in four dimensions, in which one or each of the two auxiliary scalars 
is replaced by the field strength of a gauge three-form. 
These theories are known as three-form supergravity 
and complex three-form supergravity, respectively. 
For each of them, we present a  super-Weyl invariant coupling of 
supergravity to the supermembrane and prove 
 kappa-invariance of the resulting action.
In the case of  three-form supergravity, 
we demonstrate that the action constructed reduces to that 
given by  Ovrut and Waldram twenty years ago upon imposing a super-Weyl 
gauge in which the compensating three-form superfield is set to a constant. 

\end{abstract}

\vfill

\vfill
\end{titlepage}

\newpage
\renewcommand{\thefootnote}{\arabic{footnote}}
\setcounter{footnote}{0}

\tableofcontents{}
\vspace{1cm}
\bigskip\hrule

\allowdisplaybreaks


\section{Introduction}
\setcounter{equation}{0}

The old minimal formulation for $\cN=1$ supergravity
in four dimensions, first presented in superspace \cite{WZ} and 
soon after developed in the component setting \cite{old1, old2},
is probably the most famous off-shell supergravity theory.\footnote{It 
is quite remarkable that the superspace \cite{WZ} and the component
\cite{old1,old2} formulations of old minimal supergravity were published
 in the same volume of Physics Letters B with an interval of one month.}
The field content of this theory is known to everyone   who 
studied supersymmetric field theory from the book by Wess and Bagger \cite{WB}
(part of which is  a review and extension of the approach 
pursued by Wess and Zumino \cite{WZ}).
Its physical fields are  the vielbein $e_m{}^a$ and the Majorana gravitino 
$(\psi_m{}^\a , \,\bar{\psi}_m{}_\ad )$. 
Its auxiliary fields are the vector $b_a$, the complex scalar $M$
and its conjugate $\bar M$. 
The Ferrara-van Nieuwenhuizen formulation \cite{old2}
made use of  the two real auxiliary scalars
contained  in $M = {\rm Re} \,M +\ri\, {\rm Im} \,M$.
However, the work by Stelle and West \cite{old1}
also provided a variant supergravity formulation in which 
each of the two real auxiliary scalars, ${\rm Re} \,M $ and $ {\rm Im} \,M$ 
was replaced by the field strength of a gauge three-form. 
Three years later,  Gates and Siegel \cite{GS} pointed out 
the existence of one more variant  formulation of supergravity
in which 
just one of the two real scalars in $M = {\rm Re} \,M +\ri\, {\rm Im} \,M$ 
was replaced by the field strength of a gauge three-form. 
The resulting variant formulations of
old minimal supergravity are known nowadays 
as {\it three-form supergravity} \cite{GS}
 and {\it complex three-form supergravity} \cite{old1}.\footnote{There 
 is a plausible explanation as to why Stelle and West did not describe explicitly 
the  three-form supergravity in \cite{old1}. The point is that their set of auxiliary fields
 was inspired by the structure of the gravitational vector superfield
 and its gauge freedom at the linearised level \cite{FZ1,OS,FZ2}.}
We will often refer to the  off-shell theory presented in  \cite{WZ,old2} 
 as the {\it standard formulation} or simlpy old minimal supergravity.

The difference between the standard formulation for supergravity 
and its variant realisations discussed above
can be seen from the corresponding  superfield equations of motion. 
In terms of the Grimm-Wess-Zumino superspace geometry \cite{GWZ}
(see, e.g.,  \cite{WB,Ideas} for pedagogical reviews), 
the supergravity equations corresponding to  the standard formulation were given in \cite{WZ}.
They are:
\begin{subequations}
\bea
G_a &=& 0~, \label{1.1a}\\
R &=& 0~.
\label{1.1b}
\eea
\end{subequations}
In the case of three-form supergravity, equation \eqref{1.1b} 
is replaced with 
\bea
R + \bar R =0 \quad \Longrightarrow \quad 
R - \bar R ={\rm const}~, 
\label{1.2}
\eea
while for complex three-form supergravity it turns into 
\bea
R= {\rm const}~,
\label{1.3}
\eea
in accordance with \cite{GS}. Equation \eqref{1.1a} remains the same 
for the variant formulations. Even without introducing a supersymmetric cosmological term
(which does not exist for complex three-form supergravity, see section \ref{section3.3}), 
a  negative cosmological constant is generated dynamically in the real and complex 
three-form supergravity theories for vacuum solutions 
with $R \neq 0$.\footnote{The idea that  
the use of massless gauge three-forms makes it possible to generate a cosmological constant dynamically, has attracted much interest since the early 1980s, see, e.g.,  
\cite{OS2,DvN,ANT,Hawking,Duff,BP}.} 

As is well known, every off-shell  formulation for $\cN=1$ supergravity can be realised as $\cN=1$ conformal supergravity coupled to a compensating multiplet
(see, e.g., \cite{Ideas,GGRS,FVP} for reviews).
Different off-shell formulations correspond to choosing different 
compensators. 
This leads to  another conceptual way to understand the difference between 
the standard formulation of old minimal supergravity and its two variants,
as was pointed out in \cite{GS}. In the standard formulation, the compensator 
is a {\it general}  chiral scalar superfield \cite{Siegel}. 
In Minkowski superspace, it  obeys the chirality constraint $\bar D_\ad \F =0$ and 
can be represented  as\footnote{As in \cite{WB,Ideas}, we make use of the definitions
$D^2 =D^\a D_\a$ and $\bar D^2 = \bar D_\ad \bar D^\ad$.} 
\bea
\F = -\frac{1}{4} \bar D^2 U~, 
\eea
where the prepotential $U$ is an unconstrained 
complex superfield. 
The auxiliary field of $\F$, defined by $F(x):=-\frac{1}{4}D^2\F(x,\q)|_{\q=0}$,
is a complex scalar. In the case of three-form supergravity \cite{GS}, 
the compensator is a three-form multiplet originally proposed by Gates \cite{Gates}.
It is described by a chiral superfield of the form
\bea
\Pi=-\frac{1}{4}\DB^2P~,~~~~~~
\bar{P}= P
~,
\eea
where $P$ is a real but otherwise unconstrained prepotential.
Since the prepotential $P$ is real, $\P$ is no longer a general  chiral superfield, 
for it obeys the condition 
\bea
D^2 \Pi - \bar D^2 \bar \Pi =  \ri \pa^{a} p_{a} ~, \qquad 
p_{a} =(\ts_a)^{\ad\a} [D_\a , \bar D_\ad ] P
~,
\eea
which implies that the imaginary part of the auxiliary field $F$ of $\P$
is the field strength of a gauge three-form. 
In the case of complex three-form supergravity, the compensator 
is a complex three-form multiplet proposed in \cite{GS}.\footnote{The name 
``complex three-form multiplet'' was coined in \cite{GGRS}.}
It is described by a chiral superfield of the form 
\bea
\U=-\frac{1}{4}\DB^2\bar{\S}~,~~~~~~
D^2 \bar \S=0
\label{1.7}
~.
\eea
A complex scalar $\S$ constrained by $\bar D^2 \S =0$ is called a complex linear superfield \cite{GS}.\footnote{Such a superfield was first discussed by Zumino
\cite{Zumino1980}. It is primarily used to describe the so-called non-minimal 
scalar multiplet \cite{GS}.}
Since the prepotential $\bar \S$ in \eqref{1.7} is complex antilinear, 
$\U$ is no longer a general  chiral superfield, 
for it obeys the condition 
\bea
D^2 \U  =  \ri \pa^{a} q_{a} ~, \qquad 
q_{a} =(\ts_a)^{\ad\a} [D_\a , \bar D_\ad ] \bar{\S}
~.
\eea
This property tells us that the auxiliary field 
 $F$ of $\U$  is  the field strength of a gauge {\it complex} three-form. 

Unlike the standard formulation of old minimal supergravity, 
the remarkable feature of three-form supergravity is that it 
allows a consistent coupling to the 
four-dimensional  supermembrane \cite{Achucarro:1988qb}
(the $d=4$ cousin of the $d=11$ supermembrane  \cite{Bergshoeff:1987cm,BST2})
as demonstrated by Ovrut and Waldram
 \cite{OvrutWaldram}.
 Since consistency of the supermembrane action requires the presence of
a Wess-Zumino term associated with a real gauge three-form in the target superspace
\cite{Bergshoeff:1987cm}, 
it is not surprising that this supergravity formulation plays a special role in this context.
It is natural to wonder whether
complex three-form supergravity also allows a consistent coupling 
to the supermembrane.
The main goal of this paper is indeed to show that this question has an affirmative answer.

Before we turn to the main body of the paper, 
a few   comments on the literature are in order. 
The quantum properties of a massless three-form multiplet coupled to supergravity 
were studied in \cite{BK88} (see \cite{Ideas} for a review).
The superform formulation for the three-form 
multiplet in supergravity was developed  by Bin\'etruy {\it et al.} 
\cite{Binetruy:1996xw} and used in  \cite{OvrutWaldram} to work out the 
complete component action for three-form supergravity. 
The super-Weyl invariant formulation for three-form supergravity was given in 
 \cite{KMcC}, as an extension of similar formulations for the non-minimal 
 and new minimal supergravity theories given   in section 6.6 of \cite{Ideas}. 
The formulations described in \cite{Ideas} and  \cite{KMcC} were generalised 
in \cite{FLMS} to construct the super-Weyl invariant formulation for complex three-form supergravity. Various aspects of the dynamics of three-form supergravity coupled to 
the supermembrane were studied in \cite{Bandos-Meliveo12}.

This paper is organised as follows. 
In section 2 we recall the key results concerning the formulation of $\cN=1$ conformal 
supergravity and its couplings to matter using the geometric framework of \cite{GWZ}.
Section 3 elaborates on the super-Weyl invariant formulations
for the three versions of old minimal supergravity discussed above. 
In particular, for both the real and complex three-form multiplets we provide 
a super-Weyl invariant description of the 
 gauge super 3-forms and gauge-invariant field strengths.
Section 4 describes 
consistent couplings of the real and complex 
three-form supergravity theories 
to the supermembrane.


\section{Conformal supergravity}
\setcounter{equation}{0}
\label{geometry}

As reviewed in \cite{Ideas},  
conformal supergravity can be described 
using the superspace geometry of \cite{GWZ}, 
which underlies the Wess-Zumino approach to old minimal supergravity
\cite{WZ}. Here we briefly recall the main definitions and conceptual results. 
The notation and conventions of \cite{Ideas} are used throughout this paper.

Conformal supergravity is formulated in a  curved superspace
$\cM^{4|4}$  parametrised by
local bosonic ($x^m$) and fermionic ($\q^\m, \bar \q_{\dot\m}$)
coordinates  $z^{{M}}=(x^{m},\q^{\mu},{\bar \q}_{{\dot\mu}})$,
where $m=0,1,2,3$,  $\mu=1,2$ and $\dot\mu=1,2$.
The Grassmann variables $\q^{\mu} $ and $\teb_{{\dot\mu}}$
are related to each other by complex conjugation:
$\overline{\q^{\mu}}=\teb^{{\dot\mu}}$.
We will often make use of a preferred  basis of one-forms
$E^A=(E^a,E^\a,\bar{E}_\ad)$ with dual basis $E_A=(E_a,E_\a,\bar{E}^\ad)$, 
\bea
E^A=\rd z^ME_{M}{}^A
~,~~~~~~E_A = E_A{}^M  \pa_M 
~,
\eea
which will be referred to as the supervielbein and its inverse, respectively. 
The superspace  structure group is   ${\rm SL}(2,{\mathbb{C}})$.
The covariant derivatives
have the form
\bea
\cD_{{A}}=(\cD_{{a}}, \cD_{{\a}},\cDB^\ad)
=E_{{A}}
+\O_{{A}}
~,
\label{CovDev}
\eea
where $\O_A$ stands for  the Lorentz connection, 
\bea
\O_A = \hf\,\O_A{}^{bc}  M_{bc}
= \O_A{}^{\b \g} M_{\b \g}
+\O_A{}^{\bd \gd} {\bar M}_{\bd \gd} ~,
 \label{2.3}
\eea
with $M_{bc} =-M_{cb}$,
$ M_{\b\g}=\hf(\s^{bc})_{\b\g} M_{bc}$ 
and 
${\bar M}_{\bd \gd} =-\hf(\ts^{bc})_{\bd\gd}M_{bc}$
the Lorentz generators.
These act on a covariant vector $V_c$ and 
two-component spinors $\J_\g$ and $\J_\gd$ 
as follows:
\bea
M_{ab}\,V_c=2\eta_{c[a}V_{b]}~,~~~~~~
M_{\a\b}\,\J_{\g}
=\ve_{\g(\a}\J_{\b)}~,~~~~~~
\bar{M}_{\ad\bd}\,\J_{\gd}
=\ve_{\gd(\ad}\J_{\bd)}
~.
\label{generators}
\eea

In general,  the covariant derivatives enjoy graded commutation relations of the form
\bea
{[}\cD_{{A}},\cD_{{B}}\}&=&
T_{ {A}{B} }{}^{{C}}\cD_{{C}}
+\hf R_{{A} {B}}{}^{{cd}}M_{{cd}}~,
\label{algebra-0}
\eea
where $T_{ {A}{B} }{}^{{C}}$ and $R_{{A} {B}}{}^{{cd}}$ are
the torsion and curvature tensors, respectively. 
To describe supergravity, the covariant derivatives 
have to obey certain torsion constraints \cite{WZ,GWZ} such that 
their algebra is as follows
(the expression for $[\cD_a , \cD_b]$  
is given in \cite{Ideas}):
\begin{subequations}\label{algebra}
\bea
& \{ \cD_\a , {\bar \cD}_\ad \} = -2{\rm i} \cD_{\a \ad} ~,\\
&
\{\cD_\a, \cD_\b \} = -4{\bar R} M_{\a \b}~,
 \qquad
\{ {\bar \cD}_\ad, {\bar \cD}_\bd \} =  4R {\bar M}_{\ad \bd}~, 
 \\
&\left[ \cD_{\a} , \cD_{ \b \bd } \right]
      = 
     {\rm i}  {\ve}_{\a \b}
\Big({\bar R}\,\cDB_\bd + G^\g{}_\bd \cD_\g
- \cD^\g G^\d{}_\bd  M_{\g \d}
+2{\bar W}_\bd{}^{\gd \dot{\d}}
{\bar M}_{\gd \dot{\d} }  \Big)
+ {\rm i} \cDB_\bd {\bar R} \, M_{\a \b}~,
~~~~~~\\
&\left[ { \bar \cD}_{\ad} , \cD_{ \b \bd } \right]
      =  -{\rm i}{\ve}_{\ad \bd}
\Big(R\,\cD_\b + G_\b{}^{\dot{\g}}  \cDB_{\dot{\g}}
-\cDB^\gd G_\b{}^{\dot{\d}}
{\bar M}_{\gd \dot{\d}}
+2W_\b{}^{\g \d}
M_{\g \d} \Big)
- {\rm i} \cD_\b R  {\bar M}_{\ad \bd}~.~~~~~~~ ~~
\eea
\esubeq
The torsion tensors $R$, $G_a = {\bar G}_a$ and
$W_{\a \b \g} = W_{(\a \b\g)}$ satisfy the  Bianchi identities:
\begin{subequations}
\bea
&\cDB_\ad R= 0~,~~~~~~\cDB_\ad W_{\a \b \g} = 0~,
\\
&
\cDB^\gd G_{\a \gd} = \cD_\a R~,~~~~~~
\cD^\g W_{\a \b \g} = {\rm i} \,\cD_{(\a }{}^\gd G_{\b) \gd}~.
\eea
\end{subequations}

The definition of the torsion and curvature tensors given by eq.  \eqref{algebra-0}
can be recast in the language of superforms. Starting from 
the Lorentz connection $\O_A$ defined by \eqref{2.3}, we introduce
 the connection one-form 
\bea
\O = E^C \O_C ~, \qquad 
\O V_A = \O_A{}^B V_B  = E^C \O_{CA}{}^B V_B~, 
\qquad V_A =(V_a, \J_\a, \J^\ad )~.
\label{2.8}
\eea
Then the torsion and curvature  two-forms are 
\bsubeq
\bea
T^C&:=&\hf E^B\wedge E^AT_{AB}{}^C
=-\rd E^C+E^B\wedge{\O}_B{}^C
~,
\label{torsion-def}
\\
{R}_C{}^D &:=&
\hf E^B\wedge E^AR_{AB}{}_C{}^D
=\rd{\O}_C{}^D
-{\O}_C{}^E\wedge{\O}_E{}^D
~.
\eea
\esubeq

The gauge group of conformal supergravity includes 
superspace general coordinate transformations and local Lorentz ones.
Such a transformation acts  on the covariant derivatives
and any tensor superfield $U$ (with its indices suppressed) by the rule
\begin{subequations}\label{A.5}
\bea
\d_\cK \cD_A = [\cK, \cD_A] ~, \qquad \d_\cK U = \cK U~, 
\eea
where the gauge parameter $\cK$ has  form 
\bea
\cK = \x^B \cD_B + \hf K^{bc} M_{bc} 
= \x^B \cD_B + K^{\b\g} M_{\b\g } + \bar K^{\bd  \gd } \bar M_{\bd \gd} = \bar \cK
\label{K}
\eea
\end{subequations}
and describes a coordinate transformation generated 
by the supervector field $\x=\x^B  E_B$ and a local Lorentz transformation generated 
by $K^{bc} $. 

It was first realised by Howe and Tucker  \cite{HT} that 
the algebra  \eqref{algebra} is
invariant under super-Weyl transformations
of the form 
\begin{subequations} 
\label{superweyl}
\bea
\d_\s \cD_\a &=& ( {\bar \s} - \hf \s)  \cD_\a + \cD^\b \s \, M_{\a \b}  ~, \\
\d_\s \cD_{\a\ad} &=& \hf( \s +\bar \s) \cD_{\a\ad} 
+\frac{\ri}{2} \bar \cD_\ad \bar \s \,\cD_\a + \frac{\ri}{2}  \cD_\a  \s\, \bar \cD_\ad
+ \cD^\b{}_\ad \s\, M_{\a\b} + \cD_\a{}^\bd \bar \s\, \bar M_{\ad \bd}~,
~~~~~~
\eea
\end{subequations}
accompanied by 
the following transformations of the torsion superfields
\begin{subequations} 
\bea
\d_\s R &=& 2\s R +\frac{1}{4} (\bar \cD^2 -4R ) \bar \s ~, \\
\d_\s G_{\a\ad} &=& \hf (\s +\bar \s) G_{\a\ad} +\ri \cD_{\a\ad} ( \s- \bar \s) ~, 
\label{s-WeylG}\\
\d_\s W_{\a\b\g} &=&\frac{3}{2} \s W_{\a\b\g}~.
\label{s-WeylW}
\eea
\end{subequations} 
Here the super-Weyl parameter $\s$ is a covariantly chiral scalar superfield,  $\bar \cD_\ad \s =0$.

The gauge group of conformal supergravity 
is defined to be generated by the local transformations  \eqref{A.5} and \eqref{superweyl}. It may be shown that this gauge freedom 
indeed leads to the multiplet of $\cN=1$ conformal supergravity at the component 
level (see, e.g., \cite{Ideas} for a review).

Of special importance in conformal supergravity are super-Weyl primary multiplets 
(here we follow the terminology recently used in \cite{KMT}).
A tensor superfield $T $ (with its  indices suppressed)
is said to be (super-Weyl) primary of weight $(p,q)$
if its super-Weyl transformation law is 
\bea
\d_\s T =\big(p\, \s + q\, \bar \s \big) T~,
\eea
for some  parameters $p$ and $q$.
The conformal dimension of $T$ is given by $(p+q)$.

An important class of tensor superfields
are covariantly chiral 
superfields $\F$ constrained by 
$\bar \cD_\ad \F=0$.
Due to the integrability condition $\{\bar \cD_\ad,\bar\cD_\bd\} \F=4R\bar M_{\ad \bd} \F=0$, such superfields 
 may carry only undotted indices \cite{Zumino}.
If $\F$ is covariantly chiral and  super-Weyl primary,
its weight is necessarily $(p,0)$.
We will call $\F$ a chiral primary superfield of weight $p$.

Covariantly chiral tensor superfields may be constructed using 
 the chiral projection operator \cite{WZ,Zumino}
\bea
\bar{\D}:=-\frac{1}{4}\big(\cDB^2-4R\big)
~.
\eea
Given a tensor superfield $T$ with  undotted spinor indices only, 
$\bar{\D} T$ is covariantly chiral,  $\cDB_\ad\bar{\D} T=0$.
If $T$ is a super-Weyl primary superfield of  weight $(p-2,1)$,
then $\bar{\D} T$ is a chiral primary superfield of weight $p$.
This may be checked by using the super-Weyl transformation 
of the chiral projection operator
\bea
\d_\s\bar{\D}
&=&
( 2 \s - {\bar \s})\bar\D
+\frac{1}{2}(\cDB_\ad{\bar \s})\bar \cD^\ad
+\frac{1}{4} (\bar \cD^2 \bar \s)
-\frac{1}{2} ( \bar \cD^\ad  {\bar \s} ) \cDB^\bd {\bar M}_{\ad \bd}
~,
\label{sW-chiral-proj}
\eea
which follows from \eqref{superweyl}.

Given a matter dynamical system coupled to conformal supergravity, 
its action functional must be invariant under  the local transformations  \eqref{A.5} and \eqref{superweyl}. There are two general action principles.
Given a  primary real  scalar Lagrangian $\cL =\bar \cL$ 
of weight $(1,1)$, 
the action
\bea
S&=&\int \rd^4x\rd^2\q\rd^2\qb  \,E \,\cL
~,~~~
\qquad E= {\rm Ber}(E_M{}^A)
~,
\label{N=1Ac}
\eea
is invariant under the supergravity gauge group. 
Its super-Weyl invariance follows from the transformation law 
 $\d_\s E=-(\s+\bar\s) E$.
Given a scalar chiral primary Lagrangian $\cL_{\rm c}$ of  weight $+3$, 
the  {\rm chiral} action
\bea
S_{\rm c}=\int\rd^4x\rd^2\q\, \cE \,\cL_{\rm c} 
\label{chiralAc}
\eea
is invariant under the supergravity gauge group. 
Its super-Weyl invariance follows from the transformation law $\d_\s\cE=-3\cE$
of the chiral density $\cE$. The latter may be defined in terms of a chiral prepotential
\cite{Siegel}. Alternatively, the chiral density can be read off using the general formalism of integrating out fermionic dimensions, which was developed in 
\cite{KT-M-2008-2}.\footnote{This formalism naturally leads to the appearance of 
the $\Q$ variables postulated in \cite{WB}.}
The full superspace action \eqref{N=1Ac}  can be represented 
as an integral over the chiral subspace, 
\bea
\int\rd^4x\rd^2\q\rd^2\qb\, E \,\cL
= \int\rd^4x\rd^2\q\, \cE \,
\bar\D\cL~.
\label{full-chiral}
\eea

The chiral action \eqref{chiralAc} can be represented as an integral over the full superspace, 
\bea
S_{\rm c} = \int\rd^4x\rd^2\q\rd^2\qb\, E \, {\frak C} \cL_{\rm c}~,
\eea
where ${\frak C} $ is an {\it improved complex linear} superfield\footnote{Such 
 a superfield is the conformal compensator for the 
non-minimal $\cN=1$ AdS supergravity 
\cite{BKdual}.} 
that is defined by the following properties: (i) ${\frak C} $ obeys the constraint 
\begin{subequations} \label{322}
\bea
\bar \D {\frak C} =1~;
\eea
(ii) ${\frak C} $ is super-Weyl primary of weight $(-2, 1)$, 
\bea
\d_\s {\frak C} =  (\bar \s -2 \s){\frak C} ~.
\eea
\end{subequations}
A possible choice for ${\frak C} $ is 
\bea
{\frak C}  = \frac{\bar \eta} {\bar \D \bar \eta} ~, \qquad \bar \cD_\ad \eta =0~, 
\qquad \d_\s \eta =  \s \eta~,
\eea 
for some covariantly chiral superfield $\eta$ 
such that $\D \eta $ is nowhere 
vanishing. In case ${\frak C} $ is not required to be super-Weyl primary, 
it can be identified with $R^{-1}$, 
\bea
S_{\rm c}=\int\rd^4x\rd^2\q\rd^2\qb\, E \,\frac{\cL_{\rm c}}{R}~,
\eea
provided $R$ is nowhere vanishing. This representation was discovered in 
 \cite{Siegel,Zumino}.

To conclude this section, we point out that there is an alternative 
way to define the chiral action \eqref{chiralAc}
that follows from the superform approach to 
the construction of supersymmetric invariants \cite{Castellani,Hasler,Ectoplasm,GGKS}.
It is based on the use of
the  following  super 4-form 
\bea
\Xi_4[\cL_{\rm c}]&=&
2\ri \bar{E}_\dd \wedge \bar{E}_\gd\wedge  E^b\wedge  E^a(\ts_{ab})^{\gd\dd}
\cL_{\rm c}
+\frac{\ri}{6}\ve_{abcd}\bar{E}_\dd\wedge  E^c\wedge E^b\wedge  E^a (\ts^d)^{\dd\d}\cD_\a\cL_{\rm c}
\non\\
&&
-\frac{1}{96}\ve_{abcd}E^d\wedge E^c\wedge E^b\wedge E^a 
\big(\cD^2-12\bar{R}\big)\cL_{\rm c}~,
\label{Sigma_4}
\eea
which was constructed by Bin\'etruy {\it et al.} 
\cite{Binetruy:1996xw} and independently by Gates {\it et al.} \cite{GGKS}.
This superform\footnote{In the flat-superspace 
limit, the superform \eqref{Sigma_4} reduces to the one given in \cite{Gates}.}
 is closed, 
 \bea
 \rd \, \X_4 [\cL_{\rm c}] =0~.
 \eea
 The chiral action \eqref{chiralAc} can be recast
  as an integral of $\Xi_4[\cL_{\rm c}]$ over a spacetime $\cM^4$,
 \bea
 S_{\rm c} = \int_{\cM^4} \Xi_4[\cL_{\rm c}]~,
 \label{2.24}
 \eea
where $\cM^4$ is the bosonic body of the curved superspace  $\cM^{4|4}$
obtained by switching off  the Grassmann variables. 
 It turns out that the representation \eqref{2.24} provides the simplest 
 way to reduce the action from superfields to components. 
 
Using the super-Weyl transformation law of the supervielbein
\bea
\d_\s \bar{E}_\ad&=&
(\hf\bar\s-\s)\bar{E}_\ad 
+\frac{\ri}{4} (\cD^\a \s)(\s_b)_{\a\ad}  E^b
~,~~~~~~
\d_\s E^a=
-\hf(\s+\bar\s) E^a
~,
\eea
which follows from \eqref{superweyl},
one can check that  $\X_4  [\cL_{\rm c}]$ is  super-Weyl invariant, 
\bea
\d_\s \X_4 [\cL_{\rm c}]=0~.
\label{2.26}
\eea 
This property also 
follows from the  description of this superform 
given in appendix B of \cite{BKN} where  $\X_4 [\cL_{\rm c}]$  was 
formulated in $\cN=1$ conformal superspace \cite{Butter}.
The super-Weyl invariance of $\X_4 [\cL_{\rm c}]$ 
 will be important for our subsequent analysis.


\section{Variant formulations of old minimal supergravity}
\setcounter{equation}{0}

As described in section 6.6 of \cite{Ideas}, any off-shell formulation of 
$\cN=1$ supergravity may be realised as a super-Weyl invariant coupling
of the old minimal supergravity to a conformal compensator.
Here we review the relevant realisations for the three versions of 
old minimal supergravity discussed in section 1.


\subsection{Old minimal supergravity}
\label{old-minimal-1}

In the case of the standard formulation, the conformal compensator 
is a primary chiral scalar superfield $\F$ of weight $+1$, 
\bea
\bar \cD_\ad \F =0~, 
\qquad \d_\s \F = \s \F~,
\eea
which is required to be nowhere vanishing such that $\F^{-1}$ exists.  
 The locally supersymmetric and super-Weyl invariant  action for supergravity is
\bea
S_{\text{SG,om}} = -\frac{3}{\k^2} \int {\rm d}^4x \rd^2\q\rd^2\qb
\,E\, \bar \F  \F
+ \Big\{ \frac{\m}{\k^2} \int {\rm d}^4x {\rm d}^2 \q \,\cE\,   \F^3  + {\rm c.c.} \Big\} ~,
\label{old-minimal}
 \eea
where $\k$ is the gravitational coupling constant, and 
$\m$ is a complex parameter related to the cosmological constant.
The second term in the action is the supersymmetric cosmological term
which was proposed in  
\cite{Townsend,FvN2} and then recast in the superspace setting in \cite{Siegel}.

The super-Weyl gauge freedom allows us to choose the condition $\F=1$. 
Then \eqref{old-minimal} turns into the supergravity action proposed 
in  \cite{WZ} for $\m=0$ and then generalised  to the  $\m \neq0$ case
in \cite{Siegel}. 

The equation of motion for the chiral compensator is easy to read off 
from \eqref{old-minimal}
\bea
{\mathbb R}=\m~, \qquad 
{\mathbb R}:= \F^{-2} \bar \D \bar \F
~.
\label{EOM1}
\eea
It can be shown \cite{Ideas}
that the equation of motion for the gravitational superfield can be written in the form
\bea
{\mathbb G}_{\a\ad}=0~, \qquad 
{\mathbb G}_{\a\ad} 
:=  
\Big([\cD_{\a} , \bar \cD_{\ad}]+G_{\a\ad} \Big) (\Phi \bar \Phi)^{-1/2} 
~.
\label{3.3}
\eea
The superfields $\mathbb R$ and ${\mathbb G}_{\a\ad}$ 
are super-Weyl invariant. 
Equation \eqref{3.3} is  equivalent to eq. (6.6.10) in \cite{Ideas}.
The latter states that the supercurrent of the chiral superfield $\F$,
whose dynamics is described by the action \eqref{old-minimal},
 vanishes on the mass shell. 

In the super-Weyl gauge $\F =1$, the primary superfields 
 $\mathbb R$ and ${\mathbb G}_{\a\ad}$ turn into 
the torsion superfields $R$ and $G_{\a\ad}$, respectively.  



\subsection{Three-form supergravity}

We now discuss three-form supergravity in some detail. 
First of all  we review 
the super-Weyl invariant formulation of this theory given in 
\cite{KMcC}. 
The corresponding  conformal compensator is a  three-form multiplet
coupled to conformal supergravity.
It is described by a covariantly chiral scalar $\P$ and its conjugate $\bar \P$, 
with $\P$ defined by  
\bea
\P=\bar \D
P
\qquad {\bar P} =P
~,
\label{F4-P}
\eea
where the scalar prepotential $P$ is real but otherwise unconstrained.  
The compensator $\P$ has to be nowhere vanishing so that $\P^{-1}$ exists.
We postulate  $P$ to be super-Weyl primary of  weight $(1,1)$, 
\begin{subequations}\label{s-weyl-P}
\bea
\d_\s P &=& (\s+\bar\s)  P~,
\eea
which implies that $\P$ is also primary, 
\bea
\d_\s \Pi &=& 3\s\Pi~.
\eea
\end{subequations}
As is seen from 
 \eqref{F4-P}, the prepotential $P$ is defined modulo
 gauge transformations of the form 
\bea
\d_L P = L~, 
 \qquad  \bar \D
L
=0~, \qquad \bar L =L
~,
\label{gauge-inv-P}
\eea
with the gauge parameter $L$ being a linear multiplet.\footnote{In the case of  
$\cN=1$ Poincar\'e supersymmetry, the 
linear multiplet was first introduced in \cite{FWZ}. It is used to describe 
the $\cN=1$ tensor multiplet \cite{Siegel-tensor}.}

The action for three-form supergravity is obtained from 
 \eqref{old-minimal} by replacing  $\F$ with $\Pi^{1/3}$. 
 This leads to
\bea
S_{\text{SG,t-f}} &=&-\frac{3}{\k^2} \int \rd^4x\rd^2\q\rd^2\qb
\,E\,\Big\{
\big(\bar \P \P\big)^{\frac{1}{3}}
- \hf mP
\Big\} \non \\
&=& -\frac{3}{\k^2} \int \rd^4x\rd^2\q\rd^2\qb
\,E\,
\big(\bar \P \P\big)^{\frac{1}{3}}
 + \Big\{ \frac{m}{\k^2} \int {\rm d}^4x {\rm d}^2 \q \,\cE\,   \P  + {\rm c.c.} \Big\}
~,
\label{3-form_sugra}
\eea
where $m$ is a real parameter.
By construction the action is invariant under 
gauge transformations \eqref{gauge-inv-P}.

Making use of  \eqref{F4-P},
the equation of motion for the compensator can be written as 
\bea
{\mathbb R} +\bar {\mathbb R}=2m~, \qquad 
{\mathbb R}:=  \P^{-2/3} \bar \D \bar \P^{1/3}~,
\label{3.9}
\eea
compare with \eqref{EOM1}. The chirality of ${\mathbb R}$ then implies 
\bea
{\mathbb R} =\m =\text{const}~, \qquad {\rm Re}\, \m =m~.
\label{3.455}
\eea
Unlike  \eqref{old-minimal}, 
the action \eqref{3-form_sugra}  
for three-form supergravity contains only one 
real parameter, $m$, which determines the corresponding supersymmetric cosmological term. 
However, on the mass shell 
 ${\mathbb R}$ becomes a complex  parameter, $\m$, as in \eqref{EOM1}.
 The real part of  ${\mathbb R}$ is fixed by  the equation \eqref{3.9}, 
 while its imaginary part is 
generated dynamically.
 
The three-form multiplet has a geometric realisation in terms of a gauge 
super 3-form  \cite{Binetruy:1996xw} that extends the flat-superspace 
construction of \cite{Gates}. 
Following  \cite{Binetruy:1996xw}, we consider the  real super 3-form 
\bea
R_3[P]&=&
-\ri\bar{E}_\gd\wedge E^\b\wedge E^a\,(\s_a)_\b{}^\gd P
\non\\
&&
-\hf E^\g\wedge E^b\wedge E^a  (\s_{ab})_{\g\d}\cD^{\d}P
-\hf \bar{E}_\gd\wedge E^b\wedge E^a  (\ts_{ab})^{\gd\dd}\cDB_{\dd}P
\non\\
&&
-\frac{1}{48}\ve_{abcd}E^c\wedge E^b\wedge E^a 
\big((\ts^d)^{\gd\g}{[} \cD_\g,\cDB_{\gd}{]}
+12G^d \big)P
~,
\label{B3-real}
\eea
which is constructed in terms of the prepotential $P$. 
Its exterior derivative, $R_4 :=\rd R_3$,
proves to involve $P$ only via the gauge-invariant field strength
$\Pi=\bar\D P$. For the super 4-form $R_4\equiv R_4[\Pi]$ we obtain
\bea
R_4[\Pi]
&=&
-\bar{E}_\dd \wedge \bar{E}_\gd\wedge  E^b\wedge  E^a(\tilde{\s}_{ab})^{\gd\dd}\Pi
-E^\d \wedge E^\g\wedge  E^b\wedge  E^a (\s_{ab})_{\g\d}\bar{\Pi}
\non\\
&&
-\frac{1}{12}\bar{E}_\dd\wedge  E^c\wedge E^b\wedge  E^a \ve_{abcd}(\tilde{\s}^d)^{\dd\a}\cD_\a \Pi
+\frac{1}{12}E^\d\wedge  E^c\wedge E^b\wedge  E^a \ve_{abcd}(\s^d)_{\d\ad}\cDB^\ad\bar{\Pi}
\non\\
&&
-\frac{1}{192}E^d\wedge E^c\wedge E^b\wedge E^a \ve_{abcd}\Big\{
\ri\big(\cD^2-12\bar{R}\big)\Pi
-\ri\big(\cDB^2-12R\big)\bar{\Pi}
\Big\}
~.
\label{R4}
\eea
Note that the real super 4-form $R_4$ is related to the imaginary part of the complex 
super 4-form
$\Xi_4$ in eq.~\eqref{Sigma_4} with the chiral Lagrangian $\cL_{\rm c}$ replaced 
with $\Pi$, that is
 \bea
 R_4[\Pi]=\frac{\ri}{2} \Big(\Xi_4[\Pi]-\bar{\Xi}_4[\bar{\Pi}]\Big)~.
 \eea
The field strength $R_4[\Pi]$ is invariant under gauge transformations 
of the potential $R_3[P]$ of the form
\bea
\d_L R_3[P]= R_3[L]~,
\label{3144}
\eea
where $R_3[L]$ is obtained from \eqref{B3-real} by replacing $P$ with 
a real linear superfield constrained as in  \eqref{gauge-inv-P}.
The super 3-form $R_3[L]$ coincides, modulo an overall numerical factor, 
with the field strength of the linear multiplet, see, e.g.,
\cite{BKN,Binetruy:2000zx}.

The important property of  $R_3[P]$, which was not noticed in  \cite{Binetruy:1996xw},
 is that this superform is super-Weyl invariant, 
\bea
\d_\s R_3 [P]=0 \quad \Longrightarrow \quad \d_\s R_4 [\P]=0~.
\eea

The above superform realisation of the three-form multiplet may be given 
a more geometric setting, in the spirit of \cite{Gates,Binetruy:1996xw,Binetruy:2000zx}.
This multiplet can be described by a gauge super 3-form $B_3=\frac 16 E^C\wedge E^B\wedge E^AB_{ABC}$ defined modulo gauge transformations 
\bea
\d_\L B_3 = \rd \L_2~, \qquad \L_2=\hf  E^B\wedge E^A \L_{AB}~,
\label{gauegfreedom2}
\eea
with the gauge parameter $\L_2$ being an arbitrary super 2-form.
In order to obtain an irreducible supermultiplet, the gauge invariant field strength 
$H_4 = \rd B_3$ must be subject to certain constraints such that their general solution 
is given by $H_4 = R_4[\P]$, eq.
 \eqref{R4}. Then the gauge freedom \eqref{gauegfreedom2}
may be used to choose $B_3$ in the form $B_3 = R_3[P]$, eq.
\eqref{B3-real}.
In this gauge,
the residual gauge invariance is described by \eqref{3144}.


\subsection{Complex three-form supergravity}\label{section3.3}

The super-Weyl invariant formulation for complex three-form supergravity 
was given in \cite{FLMS}.
The conformal compensator for this theory is
 a complex three-form multiplet coupled to conformal supergravity.
This multiplet is described in terms of a covariantly chiral scalar $\U$ and 
its conjugate $\bar \U$ defined as follows:
\bea
\U =
\bar \D
 \bar \S~,
 ~~~~~~
 \bar\U=\D\S
 ~.
\label{Upsilon}
\eea
Here $\S$ is a covariantly {\it complex linear} scalar superfield constrained by 
\bea 
\bar \D
\S=0
~.
\label{complex-linear}
\eea
In general, if  $\S$ is chosen to be super-Weyl primary, 
then its weight has to be $(p-2,1)$, for some $p$,
\bea
\d_\s\S=[(p-2)\s+\bar{\s}]\S
~,
\label{3188}
\eea
as a consequence of the condition that the constraint \eqref{complex-linear}
be super-Weyl invariant \cite{Ideas}.
Requiring the chiral scalar $\U=\bar\D\bar\S$ to be super-Weyl primary as well,
we have to choose  $p=3$, which means
\bea
\d_\s\S=(\s+\bar{\s})\S \quad \Longrightarrow \quad
\d_\s \U=3 \s\U~.
\label{3.18}
\eea
In order for $\U$ to be used as a conformal compensator, $\U^{-1}$ must exist.

The general solution to the constraint \eqref{complex-linear} 
is known \cite{Ideas,GGRS} to be 
\bea
\S = \cDB_\ad \bar{\J}^\ad~,
\label{3.20}
\eea
where $\bar \J^\ad $ is an unconstrained spinor superfield defined modulo 
gauge transformations 
\bea
\d_\L \bar \J^\ad = \bar \cD_\bd \bar \L^{(\ad \bd)}~,
\eea
which leave $\S$ invariant. The super-Weyl transformation of the prepotential 
can be chosen to be
\bea
\d_\s \bar \J^\ad =\frac{3}{2} \bar \s \bar \J^\ad~,
\eea
and this transformation law implies \eqref{3.18}.\footnote{More generally, 
if the super-Weyl transformation of $\S$ is given by \eqref{3188}, 
then the prepotential 
$\bar \J^\ad $ defined by \eqref{3.20} transforms as follows:
$\d_\s \bar \J^\ad =[ (p-3) \s +\frac{3}{2} \bar \s ]\bar \J^\ad$, 
as shown in \cite{Ideas}.}

The superfields $\U$ and $\bar \U$ defined by \eqref{Upsilon}
are invariant under gauge transformations of the form
\bea
\d_L \S = L_1 - \ri L_2~, \qquad \bar \D
L_i
=0~, \qquad \bar L_i =L_i
\label{3.23}
\eea
This may be recast as a gauge transformation of the prepotential 
$\bar{\J}^\ad$ defined by \eqref{3.20}, 
\bea
\d_L \bar{\J}^\ad = \bar \eta^\ad ~, \qquad \cD_\a \bar \eta^\ad=0~.
\eea

The action for complex three-form supergravity 
is obtained from \eqref{old-minimal} by replacing $\F$ with $\U^{1/3}$, which leads to
\bea
S_{\text{SG,ct-f}} = -\frac{3}{\k^2} \int {\rm d}^4x \rd^2\q\rd^2\qb
\,E\, (\bar \U  \U)^{{1}/{3}}
~.
 \eea
No contribution comes from the cosmological term in  \eqref{old-minimal} 
since the replacement 
$\F^3\to\U=\bar\D\bar\S$ and the integration rule \eqref{full-chiral} 
give a total derivative. In other words, complex three-form supergravity 
possesses no supersymmetric cosmological term. 
This is similar to the new minimal formulation for $\cN=1$ supergravity
\cite{SohniusW1,SohniusW2,SohniusW3}.
However, unlike new minimal supergravity,
a negative cosmological constant is generated dynamically
in the case of complex three-form supergravity.
Indeed, the equation of motion for the prepotential 
$\Psi_\a$, which originates in $\bar\S=\cD^\a\Psi_\a$, is 
\bea
\cD_\a {\mathbb R} =0~, \qquad 
{\mathbb R}:= \U^{-{2}/{3}} \bar \D \bar \U^{{1}/{3}}
~.
\eea
Its general solution is 
${\mathbb R} =\m =\text{const}$, where $\m$ is an arbitrary complex constant.

The complex three-form multiplet  
has a geometric superform realisation 
that extends the flat-superspace formulation of \cite{GGRS}. 
Let us consider the following complex super 3-form
\bea
C_3[\bar\S]&=&
-2\bar{E}_\gd\wedge E^\b\wedge E^a\,(\s_a)_\b{}^\gd\bar\S
\non\\
&&
+\ri E^\g\wedge E^b\wedge E^a  (\s_{ab})_{\g\d}\cD^{\d}\bar\S
+\ri \bar{E}_\gd\wedge E^b\wedge E^a  (\ts_{ab})^{\gd\dd}\cDB_{\dd}\bar\S
\non\\
&&
+\frac{\ri}{24}\ve_{abcd}E^c\wedge E^b\wedge E^a 
\big((\ts^d)^{\gd\g}{[} \cD_\g,\cDB_{\gd}{]}
+12G^d \big)\bar\S
~.
\label{B3-complex}
\eea
Its exterior derivative, $\rd C_3[\bar\S]$,
proves to be constructed 
entirely in terms of the field strength
$\U=\bar\D\bar\S$. More precisely, it holds that 
\bea
\rd C_3[\bar\S]
=\Xi_4[\U]~,
\eea
where $\Xi_4[\U]$ is the complex super 4-form in eq.~\eqref{Sigma_4} 
with $\cL_{\rm c}$ replaced by $\U$.
Similar to the super 3-form \eqref{B3-real}, 
$C_3[\bar\S]$ 
is super-Weyl invariant, 
\bea
\d_\s C_3[\bar\S] =0 \quad \Longrightarrow \quad \d_\s \Xi_4[\U]=0
~.
\eea

The field strength $\X_4[\U]$ is invariant under gauge transformations 
of the potential $C_3[\bar \S]$ of the form
\bea
\d_L C_3[\bar \S]= C_3[L_1 +\ri L_2]~,
\eea
where $C_3[L_1 +\ri L_2]$ is obtained from \eqref{B3-complex}
by replacing $\bar \S \to L_1 +\ri L_2$, with 
the gauge parameters $L_i$ constrained as in \eqref{3.23}.


\section{Supermembrane coupled to supergravity}

We are now in a position to formulate consistent dynamics of a supermembrane 
propagating in  a three-form supergravity background. Our construction 
will be  valid for both the real and complex three-form supergravity theories. 
We will draw heavily on the results of  \cite{OvrutWaldram,Bergshoeff:1987cm}.

The action for a supermembrane propagating in a  three-form
 supergravity background is proposed to be
\bea
S=T_3\int\rd^3\x\,\Big\{
\hf\sqrt{-\g}\g^{ij}\F\bar\F E_i{}^aE_j{}^b\eta_{ab}
+\frac{1}{6}\e^{ijk}E_i{}^CE_j{}^BE_k{}^AB_{ABC}
-\hf \sqrt{-\g}\Big\}
~.
\label{membrane}
\eea
Here $\x^i$, with $i=1,2,3$, are the  coordinates of the world volume
with metric $\g_{ij}$, 
$\g=\det (\g_{ij})=\frac{1}{6}\e^{ijk}\e^{i'j'k'}\g_{ii'}\g_{jj'}\g_{kk'}$,
 and the Levi-Civita symbol $\e^{ijk}$ is normalised as $\e^{123}=1$.
As usual,  $\g^{ij}$ denotes the inverse metric
such that $\g^{ik}\g_{kj}=\d^i_j$.
In \eqref{membrane} we have used the notation
\bea
E_i{}^A=\pa_i z^M(\x) \, E_{M}{}^A
 \label{4.2}
\eea
for the pull-back supervielbein.

Our action \eqref{membrane} involves 
a composite dilaton $\F\bar\F $, 
where $\F$ is a  chiral primary superfield of weight $+1$ such that $\F^{-1}$ exists.
The superfield $\F$ is assumed to be the compensator of one of the two  three-form supergravity theories.
In the case of three-form supergravity, we choose  $\F$ to be $\Pi^{1/3}$. 
The presence of $\F$ in \eqref{membrane}  distinguishes our action from that considered 
in  \cite{OvrutWaldram}.
In the case of complex three-form supergravity,  $\F=\U^{1/3}$.
The inclusion of the dilaton
 is necessary since we are working with the  super-Weyl invariant formulation for  supergravity.
The super-Weyl freedom may be fixed by choosing the condition $\F=1$.

The Wess-Zumino term in \eqref{membrane} involves the components of 
a gauge super 3-form $B_3=\frac 16 E^C\wedge E^B\wedge E^AB_{ABC}$. 
This superform 
is chosen as follows:
(i) for three-form supergravity, 
$B_3 = R_3[P]$,  with $R_3[P]$ defined by eq. \eqref{B3-real};
and 
(ii)  in the case of 
complex three-form supergravity,
$B_3= \frac{\ri}{2}(C_3[\bar\S]-\bar{C}_3[\S])$, 
with $C_3[\bar\S]$ given by \eqref{B3-complex}.
The latter super 3-form, $B_3[\S,\bar\S]$, 
turns out to coincide with 
the superform $R_3[\S+\bar{\S}]$, which is obtained from \eqref{B3-real} 
by replacing $P$ with $(\S+\bar{\S})$.
In both cases, the gauge-invariant field strength  $H_{4}=\rd B_3$ is such that
\bea
H_4
=\frac{\ri}{2}\big(\Xi_4[\F^3]-\bar{\Xi}_4[\Fb^3]\big)
~,
\label{H4}
\eea
where $\Xi_4$ is the superform in eq.~\eqref{Sigma_4} with  $\cL_c$ replaced 
either with $\F^3=\Pi$ or $\F^3=\U$.

Consistent supermembrane actions must possess 
a local fermionic $\k$-symmetry
\cite{Bergshoeff:1987cm}.\footnote{The $\k$-symmetry was first discovered 
in the cases of  massive \cite{deAL1,deAL2} and 
massless  \cite{Siegel:1983hh} superparticles. See \cite{Sezgin:1993xg,Sorokin}
and references therein for reviews of various aspects of the $\kappa$-symmetry .}
This gauge symmetry ensures that half of the fermionic degrees of freedom 
can be gauged away and that 
spacetime and world-volume supersymmetry can be linked to each other.
Let us now show that the action \eqref{membrane} is consistently 
$\k$-symmetric in arbitrary three-form supergravity backgrounds.

Defining $\d_\k E^A:= \d_\k z^M E_M{}^A$,
we consider the 
fermionic gauge
transformation
\bsubeq\label{ksymm000}
\bea
\d_\k E^a=0
~,
~~~~~~~~~
\d_\k E^\a=\F^{1/2}\bar{\F}^{-1}\Big(\k^\b+\bar{\k}_{\ad}\bar{\G}^{\ad\a}\Big)
~,
~~~~~~
\overline{\k^{\a}}\equiv \bar{\k}^{\ad}
~,
\label{ksymm001}
\eea
where the gauge parameter $\k^\a (\x)$ is a two-component undotted
  ${\rm SL}(2,{\mathbb{C}})$ spinor, 
and a world-volume scalar. The variation 
$\d_\k \bar{E}_\ad$ is the complex conjugate of $\d_\k E^\a$,
while  $\G_{\a\ad}$ and $\bar{\G}^{\ad\a}=-\ve^{\a\b}\ve^{\ad\bd}\,\overline{\G_{\b\bd}}$
are given by
\bea
\G_{\a\ad}
&=&
-\frac{1}{6\sqrt{-\g}}(\F\bar\F)^{{3}/{2}}\,\e^{ijk}E_i{}^{a}E_j{}^{b}E_k{}^c\ve_{abcd}(\s^{d})_{\a\ad}
~,
\\
\bar{\G}^{\ad\a}
&=&
\frac{1}{6\sqrt{-\g}}(\F\Fb)^{{3}/{2}}\,\e^{ijk}E_i{}^{a}E_j{}^{b}E_k{}^c\ve_{abcd}(\ts^{d})^{\ad\a}
~.
\eea
\esubeq
Following \cite{Bergshoeff:1987cm}, we parametrise the variation of the membrane's metric as
\bea
\d_\k\g_{ij}=2(X_{ij}-\g_{ij}X_k{}^k)
~,
\eea
with $X_{ij}$ to be determined below.
We now point out  the relation
\bea
\d_\k E_i{}^A
&=&
\pa_i \d_\k E^A
-2\d_\k E^CE_i{}^B\O_{[BC)}{}^A
+\d_\k E^CE_i{}^BT_{BC}{}^A
~,
\eea
where the Lorentz connection $\O_{BC}{}^A$ and the torsion tensor
$T_{BC}{}^A$ are given by eqs. \eqref{2.8} and
\eqref{torsion-def}, respectively. 
In conjunction with 
integration by parts,  
this relation may be used to bring the variation of the action to the form:
\bea
\d_\k S
&=&
T_3\int\rd^3\x\Big\{
-\sqrt{-\g}X^{ij}(T_{ij}-\g_{ij})
+\frac{1}{2}\sqrt{-\g}\g^{ij}T_{ij}\big(\Fb\d_\k E^\a\cD_\a\F+\F\d_\k \bar{E}_\ad\cDB^\ad\Fb\big)
\non\\
&&~~~
-\sqrt{-\g}\g^{ij}(\F\Fb)E_i{}^D\d_\k E^CT_{CD}{}^aE_j{}^b \eta_{ab}
+\frac{1}{6}\e^{ijk}E_i{}^DE_j{}^CE_k{}^B\d_\k E^AH_{ABCD}
\Big\}
~.~~~~~~
\label{var-1}
\eea
Here we have denoted 
$T_{ij}:=\F\Fb \,E_i{}^aE_j{}^b\eta_{ab}$,
and $H_{ABCD}$ represents the components of the 
closed super 4-form 
\bea
H_4=\rd B_3=\frac{1}{4!}E^D\wedge E^C\wedge E^B\wedge E^A\Big\{4\cD_{[A}B_{BCD)}
-6T_{[AB}{}^EB_{|E|CD)}\Big\}
~.
\eea

Since $\d_\k E^a=0$, in accordance with eq.~\eqref{ksymm001},
 only dimension-0 and dimension-1/2 components of the torsion tensor 
 appear  in the $\k$-variation  \eqref{var-1}.
In the case of the superspace geometry of section \ref{geometry},
no dimension-1/2 torsion is present, and 
the only dimension-0  torsion is
\bea
T_{\a}{}^{\bd}{}^c=-2\ri(\s^c)_\a{}^{\bd}
~.
\label{torsion-0}
\eea
The non-trivial components of the superform $H_4$ defined by \eqref{H4},
which appear in the variation of the Wess-Zumino term in eq.~\eqref{membrane}, are 
\bea
H_{ab}{}_{\g\d}=-4(\s_{ab})_{\g\d}\Fb^3
~,~~~~~~
H_{abc\d}=\hf\ve_{abcd} (\s^d)_{\d\dd} \bar \cD^\dd \Fb^3
~,
\label{H-0}
\eea
and 
their complex conjugates.
Now, if me make use of the relations \eqref{ksymm000},
\eqref{torsion-0} and \eqref{H-0}, 
and also choose
\bea
X^{ij}
=&\,
\Big\{&-\frac{2}{\sqrt{-\g}}\,\F^{1/2}\Fb^{2}\,
\e^{kl(i}\g^{j)p}E_{k}{}^{b}E_{l}{}^{c}E_p{}^\a (\s_{bc})_{\a\b}
\non\\
&&
-6\ri\F^{3/2}(T_{[k}{}^{k}+\d_{[k}^k)\d_l^{(i}E_{p]}{}_\bd\g^{j)l}\g^{pq}E_{q}{}^a(\s_{a})_\b{}^\bd
\non\\
&&
+\frac{1}{12\sqrt{-\g}}\,\F^{1/2}\Fb\,
\e^{klp}\g^{ij}E_k{}^{a}E_l{}^{b}E_p{}^c\ve_{abcd}(\s^{d})_{\b\bd}\cDB^\bd\Fb
\non\\
&&
+\frac{3}{2}\F^{-1/2}\Fb^{-1}(T_{[k}{}^{k}+\d_{[k}^k)T_{l}{}^{l}\d_{p]}^{(i}\g^{j)p}\cD_\b \F
\Big\}{\k}^{\b}
+{\rm c.c.}
~,
\eea
it may be shown, using some algebra,  that the action 
\eqref{membrane} is indeed invariant
under the  fermionic gauge transformation.

We emphasise that the Wess-Zumino term in the supermembrane action  \eqref{membrane}
is constructed in terms of the gauge three-form, 
for which our results in eqs.~\eqref{B3-real} and 
\eqref{B3-complex} are essential.
On the other hand, 
the proof of $\k$-invariance, which was first given in \cite{OvrutWaldram} in the gauge $\F=1$,
requires only the constraints on the torsion and the field strength four-form, 
and for this reason is blind to the concrete three-form supergravity we choose, real or complex.

\section{Concluding comments}

The super-Weyl invariant formulation for the real and complex 
three-form supergravity theories
provides a simple description for conformally flat supergravity backgrounds
(compare with section 6.5 of \cite{Ideas}).  
It is obtained by choosing
the supergravity covariant derivatives $\cD_A$ to coincide with the flat global ones $D_A=(\pa_a , D_\a, \bar D^\ad)$, while keeping the corresponding compensator, $\P$ or $\U$,  to be arbitrary.  
In the case of three-form supergravity, our action \eqref{membrane}
then reduces to that describing the supermembrane coupled to 
a background three-form multiplet, 
as constructed by Bandos and Meliveo \cite{Bandos-Meliveo}. 

Since all supersymmetric actions described in this paper are super-Weyl invariant, 
 our results may be recast in the framework of 4D $\cN=1$ conformal superspace
\cite{Butter}.

Recently, nilpotent three-form multiplets have been used to construct theories for 
spontaneously broken local $\cN=1$ supersymmetry \cite{FKRR,BK17}.
One can also consider Goldstino models described by a nilpotent 
complex three-form multiplet. Its off-shell structure is still given by 
the relations \eqref{Upsilon}--\eqref{3.18}. But now it is subject to the nilpotency 
constraint 
\bea
\U^2=0~,
\eea
 with the additional condition that $\D \U$ is nowhere vanishing such that 
 $(\D \U)^{-1}$ exists. This Goldstino superfield may be coupled to every off-shell 
 formulation for supergravity. In particular, its coupling to old minimal supergravity 
 is described by the action
 \bea
S_{\text{Goldstino}} = \int {\rm d}^4x \rd^2\q\rd^2\qb
\,E\, \frac{\bar \U  \U}{(\bar \F \F)^2}~.
 \eea

~~~\\
\noindent
{\bf Acknowledgements:}\\
We are grateful to Ian McArthur, Joseph Novak 
and especially Dmitri Sorokin for comments on the manuscript
and suggestions on the literature.
GT-M is grateful to D.\,Sorokin for useful discussions. 
The work of SMK is supported in part by the Australian 
Research Council, project No.\,DP160103633.
The work of GT-M was supported by
the Interuniversity Attraction Poles Programme initiated by the Belgian Science Policy (P7/37) 
and the C16/16/005 grant of the KULeuven.


\begin{footnotesize}

\end{footnotesize}

\end{document}